\documentstyle[11pt,psfig]{article}

\newcommand\bg{\begin{eqnarray}}
\newcommand\ed{\end{eqnarray}}
\def\ra{\rightarrow}
\def\vs{\vspace{0.15in}}
\def\hs{\hspace{0.1in}}

\def\D{\partial}
\def\no{\nonumber}
\def\slashD{{/\kern-0.7em\partial}}

\begin{document}

\begin{titlepage}
\title{\bf{
Source Galerkin Calculations in \\ Scalar Field Theory}}
\author{
John W. Lawson {\dag}
{\thanks{Present Address:~ICTP, 34014 Trieste, Italy}}
and
G.S.~Guralnik
{\thanks{Research supported in part by DOE Grant DE-FG02-91ER40688 - Task D}}
\\
Department of Physics \\
Brown University \\
Providence RI 02912}
\date{}
\maketitle
\begin{abstract}
\noindent
In this paper, we extend previous work on scalar $\phi^4$
theory using the Source Galerkin method.
This approach is based on finding solutions $Z[J]$ to the lattice
functional equations for field theories in the presence of
an external source $J$.
Using polynomial expansions for the generating functional $Z$,
we calculate propagators
and mass-gaps for a number of systems.  These calculations
are straightforward to perform and
are executed rapidly compared
to Monte Carlo.  The bulk of the computation involves a single
matrix inversion.
The use of polynomial expansions illustrates in a
clear and simple way the ideas of the Source Galerkin method.
But at the same time, this choice
has serious limitations.  Even after exploiting symmetries, the size
of calculations become prohibitive except for small systems.
The calculations in this paper were made on a workstation of modest
power using a fourth order polynomial expansion
for lattices of size $8^2$,$4^3$,$2^4$ in $2D$, $3D$, and $4D$.
In addition, we present an alternative to the Galerkin
procedure that results in sparse matrices to invert.
\end{abstract}
\thispagestyle{empty}
\vskip-20cm
\noindent
\phantom{bla}
\hfill{{\bf{BROWN-HET-966}}} \\
\end{titlepage}

\section{Introduction}

Previously, we presented a new numerical approach to quantum
field theory called the Source Galerkin
method \cite{galerkin}.
This deterministic approach is fundamentally
different from other attempts to solve lattice theories.
It is based on finding solutions to the lattice
functional equations for field theories in the presence of
an external source.  The object of calculations is a generating
function $Z$ which is a multivariate function of the discretized
sources $J_i$.  There is one source variable for
each site of
the lattice.  Once an approximate solution for $Z$ is found,
the Green's functions can be extracted by differentiation.
This formulation is identical for fermions,
the only difference being
that the fermionic sources anticommute \cite{fermion}.

Despite the large number of coupled differential
equations for $Z$, one for each
source variable, there is nothing unusual about this problem
mathematically.
For scalar $\phi^4$,
the functional relations are a set of third order {\em linear}
differential equations.
We can solve them as we do any set of linear differential equations.
Since at the end of any calculation the sources are set to zero,
the obvious first choice is to expand $Z$ in a power series in
the sources about the origin.
Since we cannot solve the theory exactly,
we must introduce an approximation.
For the purposes of this paper, we
truncate the series expansion for $Z$
after some maximum power in the source variables.
In this way, we obtain a closed, finite set of linear equations
for the expansion coefficients accurate
to within the limits of the approximation.

This approach has many advantages because of the simple structure
of Taylor series expansions.
However, the computational complexity of this particular expansion
scheme increases rapidly
with system size.
As a result, truncated series can only be used for small systems.
Nevertheless, due to the accuracy of this method, it is possible
to extract large amounts of information about the system
with only short computer runs.
It should be emphasized that this shortcoming
does not represent an intrinsic limit
of the Source Galerkin method,
but rather is a reflection of choosing
power series basis functions.
Using this basis we are able to display many of the
features that make the Galerkin method so powerful
for controlling the error
resulting from approximations.
Because Taylor expansions are so simple,
we are able to present another
interesting method to deal with truncated series expansions.
Elsewere, we develop more appropriate expansion functions and
more general numerical procedures.
These are necessary for Galerkin based examinations of larger systems.

The next sections are organized as follows.
In Section I, we give a review
of the Source Galerkin method.  In Section II, we work out a simple
example for a two site model to illustrate the method and expose its
structure.  Next, we find symmetry to be a valuable tool for reducing
the size of calculations.  By constructing the power series to reflect
explicitly the lattice symmetries, we significantly reduce
the number of unknowns in our problem.  In Section III, we give
a systematic method to construct lattice invariant polynomials.
These polynomials will form a basis for our solutions.
In Section IV, we present results for numerical calculations
on systems for scalar $\phi^4$ theory in $D=2,3,4$.
Comparisons
with Monte Carlo simulations are shown.  Finally, in Sections V,VI,
we present an alternative to the Galerkin method applied to Taylor series
test functions.  Using this method we introduce
a flatness criterion to control the convergence of approximations.
This approach does not solve the problem of increasing
computational complexity, but does result in sparse matrices to invert
at the end of calculations.

\section{Source Galerkin Method}

We study quantum field theory in the
presence of an external source.
The vacuum persistence amplitude
$Z[J] = _{J} \langle 0|0 \rangle_{J}$
is the generating functional for the Green's functions.  It
is constrained by a functional differential equation.
For a self-interacting scalar field theory,
the dynamics are described by
\bg
L_{J}\,(\phi) \; = \;
1/2 \, \hat{\phi} \; (\Box + M^{2}) \; \hat{\phi} \; +
\; g/4 \, \hat{\phi^{4}} \; + \; J \hat{\phi}.
\ed
Taking vacuum expectation values of the
operator equations of motion
\bg
( \Box + M^{2} ) \langle \hat{\phi} \rangle _{J} \; + \;
g \,  \langle \hat{\phi^{3}} \rangle_{J} \;  =  \;
J \,  _{J} \langle 0  | 0 \rangle_{J}
\ed
and identifying Euclidean space expectation values
of fields with functional derivatives
of $Z[J]$
\bg
G(x_{1} \ldots x_{n}) \; = \;
\frac{\delta^{n} Z[J]}{\delta J(x_{1}) \ldots \delta J(x_n)}
\ed
yields the functional relation
\bg
( \Box + M^{2} ) \, \frac{\delta Z}{\delta J(x)}
\; + \; g \, \frac{\delta^{3} Z}{\delta J(x)^{3}}  \;  =  \; J \;Z.
\ed
After solving this equation for $Z$, we can extract all
information about our field theory by functional differentiation.
On a $D$-dimensional Euclidean lattice,
the functional equation
becomes a set of coupled differential equations
\begin{equation}
(2\, D +M^{2}){\D Z \over \D {J(i)}}
 -  \sum_{nn} {\D Z \over \D {J(i)}}
 +  g {{\D^3 Z  }\over {\D {J(i)^3}}}
 =  J(i) Z
\end{equation}
where the sum is over nearest neighbors. There
is one equation per site.

For finite lattices with $N$ sites, we construct
$Z$ as a multivariate function of the $N$
source variables $J(i)$.  To construct a solution
to the differential equations, we can expand $Z$
on any complete set of functions in the source
variables
\bg
Z \hs = \hs \sum_n \; a_n \, \phi_n (\{J\})
\ed
where the $a_n$ are the unknowns of the problem.
A particularly simple choice is polynomial functions.
It should be emphasized that other choices are possible,
and indeed preferred, for more complicated systems.
In this paper, we consider only polynomial expansions.

The number of independent unknown
coefficients in our expansion can be reduced
by exploiting the symmetries of the lattice.
$Z$ is constructed to be invariant under
the symmetry group of the lattice
\bg
Z \hs = \hs \sum_{n,m} \; a_{n,m} \, P_{n,m}
\ed
where  $P_{n,m}$  are invariant polynomials
in the source variables of order $n$
and with $m$ invariant classes for a given order.  The
coefficients $a_{n,m}$ are to be determined.  The number of
invariant classes for a given order depends on both the number
of lattice sites and on the number of symmetry operations for a
given lattice.  For example, higher dimensional lattices have
larger symmetry groups, and therefore have fewer independent unknowns.
They form a lattice invariant basis upon which we construct
solutions for $Z$.

In order to solve the differential equations,
we must specify boundary conditions.
We normalize
the vacuum amplitude
\bg
Z[J=0]  = 1
\ed
and we require odd Green's functions vanish
\bg
\frac{dZ}{dJ(i)}  |_{J=0} =  \langle \phi(i) \rangle = 0.
\ed
Rather than specify the second derivatives of $Z$,
the two-point functions,
we set the third boundary condition by truncating
$Z$ at some finite order.
The truncation order is dictated by
computational constraints.
As a boundary condition, truncation guarantees that as $g\ra0$ in
the interacting theory, the solution reduces to the free field.  This is
possible only for path integral formulations constrained
to real integration contours \cite{garcia,garciaguralnik,cooper}.
Truncation in this sense not
only sets a boundary condition, but also introduces an
approximation scheme.  This scheme
is made systematic by truncating at successively higher orders.

In order to fit our truncated solution $Z_{T}$ to a solution
of the differential equations, we use the Galerkin method \cite{fletcher}.
This is a spectral method well-known in applied mathematics,
and is especially well-suited to our problem.
In this approach, given a differential equation
\bg
\hat{D} \,Z [\,J \, ] = 0,
\ed
we can define a residual function $R$
\bg
R = \hat{D}_{i} \,Z_{T}
\ed
where $\hat{D}_{i}$ is the differential source operator
centered at site $i$.
If $Z_{T}$ were the exact solution, then $R$ would be identically
zero.  But since $Z_{T}$ is approximate, it is necessarily a
nontrivial function of the source variables.  The residual
is a direct measure of the error due to approximation.
We will fit our solution by minimizing $R$.
To implement this, we define an inner product in the source space
\begin{equation}
(g,f) = \int_{-\epsilon}^{\epsilon}
g(J(1) \ldots J(N))f(J(1) \ldots J(N))  \, [dJ]
\end{equation}
where the integration is over all $J(i)$, and $\epsilon$ is considered small.
Requiring the inner product

\clearpage

\noindent
of $R$ with linearly independent
test functions $T_{k}$ to vanish
\begin{eqnarray}
(R,T_{1}) & = & 0 \nonumber \\
\vdots & &   \\
(R,T_{j}) & = & 0 \nonumber
\end{eqnarray}
generates a set of linear algebraic equations.
We can
generate as many of these equations as we like as long as the
$T_{k}$ are linearly independent.

In analogy with a variational principle,
we chose the test functions as
\bg
T_{k} \hs = \hs \frac{d}{dJ(1)} \; P_{n,m}
\ed
where the $P_{n,m}$ are the invariant polynomial that formed the basis
for $Z_{T}$ .  This choice guarantees that we always have as many
equations as unknowns and experience has shown that it gives rapid
convergence of $Z_{T}$ to the exact solution.
We construct as many equations
as there are independent unknowns
in our power series solution
These equations can be solved by a single matrix inversion where
the resulting coefficients are the lattice Green's functions.

\section{Galerkin Solutions for Two Sites}

Zero-dimensional systems are described by a single
ordinary differential equation.  A truncated solution can be found
by simply inverting the recursion relations.  In this case,
no Galerkin procedure is needed.
But for more than one site,
we find a fundamental difficulty; namely, the
truncated equations for the expansion coefficients
are overdetermined.  These equations
can not be solved by simple inversion, and in fact, cannot
be solved exactly at all.  This is because
even though $Z$ is required to reflect all the symmetries of
our system, the individual differential source operators
$\hat{D}_{i}$ are invariant under only a subgroup, namely
reflections about site $i$.  This symmetry mismatch
causes the truncated recursion relations for the
$Z$ to be inconsistent.
It is here that the Galerkin method
is needed.  It allows us not only to construct sequences of
approximations that converge to the exact solution of the differential
equations, but also resolves, at least approximately,
the problem of overdetermined equations.

We will examine in detail the two site model to see how this scheme works.
This example illustrates all the structure needed for large scale
calculations.
We have field operators
$\phi_{1}$, $\phi_{2}$ and source variables $J_{1}$,$J_{2}$
for the two sites.
The two coupled linear differential equations are
\bg
(2+M^2) \frac{d Z}{d J_{1}}
- 2 \; \frac{d Z}{d J_{2}}
+ g \; \frac{d^{3} Z}{d J_{1}^{3}}
= J_{1} Z
\ed
\bg
(2+M^2)  \frac{d Z}{d J_{2}}
-  2 \frac{d Z}{d J_{1}}
+  g \frac{d^{3} Z}{d J_{2}^{3}}
= J_{2} Z
\ed
and we solve them with the following power series expansion
truncated after $4th$ order
\bg
Z_{T} & = & 1  +  \frac{1}{2} a_{1} (J_{1}^{2} + J_{2}^{2})
 + a_{2} J_{1} J_{2} \no \\
&& + \frac{1}{24} a_{3} (J_{1}^{4} + J_{2}^{4})
   + \frac{1}{6}  a_{4} (J_{1}^{3} J_{2} + J_{1} J_{2}^{3})
   + \frac{1}{4}  a_{5} J_{1}^{2} J_{2}^{2}.
\ed

Since the two differential equations differ only by lattice
symmetry operations, and since $Z$ has been explicitly
constructed as lattice symmetric, it is sufficient to consider
only one differential equation to constrain the coefficients
of $Z$. We choose the one centered at site 1.
After applying this equation to $Z$ and equating like
powers of $J_{i}$, we are left with 6 algebraic relations between
the 5 coefficients.  These are the lattice
Schwinger-Dyson (SD) equations.
\bg
\alpha \, a_{1} - 2 a_{2} + g \,  a_{3} - 1 & = & 0  \no \\
\alpha \, a_{2} - 2 a_{1} + g \, a_{4} & = & 0 \no \\
\alpha \, a_{3} - 2 a_{4} - 3 a_{1} & = & 0 \\
\alpha \, a_{4} - 2 a_{5} - 2 a_{2} & = & 0 \no \\
\alpha \, a_{4} - 2 a_{3} & = & 0 \no \\
\alpha \, a_{5} - 2 a_{4} - a_{1} & = & 0 \no
\ed
where the coefficients $a_i$
\bg
a_{1} & = & \langle \phi_{1}^{2} \rangle
= \langle \phi_{2}^{2} \rangle \no \\
a_{2} & = & \langle \phi_{1} \phi_{2} \rangle  \no  \\
a_{3} & = & \langle \phi_{1}^{4} \rangle
= \langle \phi_{2}^{4} \rangle \\
a_{4} & = & \langle \phi_{1}^{3} \phi_{2} \rangle
= \langle \phi_{1} \phi_{2}^{3} \rangle \no \\
a_{5} & = & \langle \phi_{1}^{2} \phi_{2}^{2} \rangle \no
\ed
are expectation values of fields and $\alpha = 2+M^2$.

To implement the Galerkin procedure for this model,
we define the following inner product
\bg
(\,g,f\,) = \int_{-\epsilon}^{\epsilon} \int_{-\epsilon}^{\epsilon}
g(J_{1}, J_{2}) f(J_{1}, J_{2}) dJ_{1} dJ_{2}
\ed
and use the following test functions
\bg
T_{1} = J_{1}, \; \;
T_{2} = J_{2}, \; \;
T_{3} = 1/6 \, J_{1}^{3}, \no \\
T_{4} = 1/2 \, J_{1}^{2} J_{2} +  1/6 \, J_{2}^{3}, \; \;
T_{5} = 1/2 \, J_{1} J_{2}^{2}
\ed
Forcing their inner products with $R = \hat{D}_{i=1} Z_{T}$ to
zero yields five equations for our five
unknown coefficients.
\bg
(\;R\;,\;T_{1}\;) & = & 0 \no  \\
(\;R\;,\;T_{2}\;) & = & 0 \no  \\
(\;R\;,\;T_{3}\;) & = & 0  \\
(\;R\;,\;T_{4}\;) & = & 0 \no \\
(\;R\;,\;T_{5}\;) & = & 0  \no
\ed
Solving these algebraic equations and taking the limit $\epsilon \ra
 0$, gives
\bg
a_{1} & = & (M^{2}+2)(19M^{4}+76M^{2}+35G) \no \\
a_{2} & = & 2\,(19M^{4}+76M^{2}-22G) \no \\
a_{3} & = & 3\,(19M^{4}+76M^{2}+35G+76) \\
a_{4} & = & 114\,(M^{2}+2) \no \\
a_{5} & = & (19M^{4}+76M^{2}+35G+228)  \no
\ed
where the $a_{i}$ are all divided by the factor
\bg
N \,=\, 19\, M^{8}\,+\,152\,M^{6}\,+\,92\,G\,M^{4}\,+\,304\,M^{4}
\,+\,368\,G\,M^{2}\,+\,105\,G^{2}\,+\,456\,G
\ed
Notice that these coefficients reduce to
the free field values as $g \ra 0$.
The free values can be obtained more directly by looking at the inverse
of the propagator matrix.
Recall that the free field limit
is important in setting our third boundary condition, and is
guaranteed by truncation.  We see that our interacting solution
for $Z$ respects this limit.

To examine the physical content of this procedure, we set
$M=1.0,g=0.5$ and let $\epsilon$ be a free variable.  The
coefficients are now rational functions of $\epsilon$.  Since
we consider $\epsilon$ small, we keep only the leading order
terms.  Thus the coefficients are
\bg
a_{1} & = &  0.3500 + 0.0009 \, \epsilon^{2} \no \\
a_{2} & = &  0.1729 + 0.0020 \, \epsilon^{2} \no \\
a_{3} & = &  0.5916 + 0.0026 \, \epsilon^{2} \\
a_{4} & = &  0.3623 + 0.0025 \, \epsilon^{2} \no \\
a_{5} & = &  0.3582 + 0.0020 \, \epsilon^{2} \no
\ed
Putting these values into the truncated SD equations yields
\bg
\alpha \, a_{1} - 2 a_{2} + g \,  a_{3} - 1 & = &  0 \no \\
\alpha \, a_{2} - 2 a_{1} + g \, a_{4} & = & 0 \no \\
\alpha \, a_{3} - 2 a_{4} - 3 a_{1} & = & 0  \\
\alpha \, a_{4} - 2 a_{5} - 2 a_{2} & = & 0.0247  \no \\
\alpha \, a_{4} - 2 a_{3} & = & - 0.0961  \no \\
\alpha \, a_{5} - 2 a_{4} - a_{1} & = & 0  \no
\ed

We see that after $\epsilon \ra 0$, the low order equations are exactly
satisfied, but some of the higher equations have nonzero right hand
sides.  They are satisfied only approximately.  This is to be expected
since the full set of SD equations is overdetermined and cannot
be solved exactly.  In the figure, we plot the partition function $Z$
using the results of eq (20)
as a function of the source variables.  The Green's functions
correspond to partial derivatives of $Z$ about the origin.  In
particular, the 2-pt functions correspond to the curvatures
of $Z$ about $J=0$.

Let us look at exactly what the Galerkin procedure has done.  Each
time we take the

\begin{figure}
\centerline{\psfig{figure=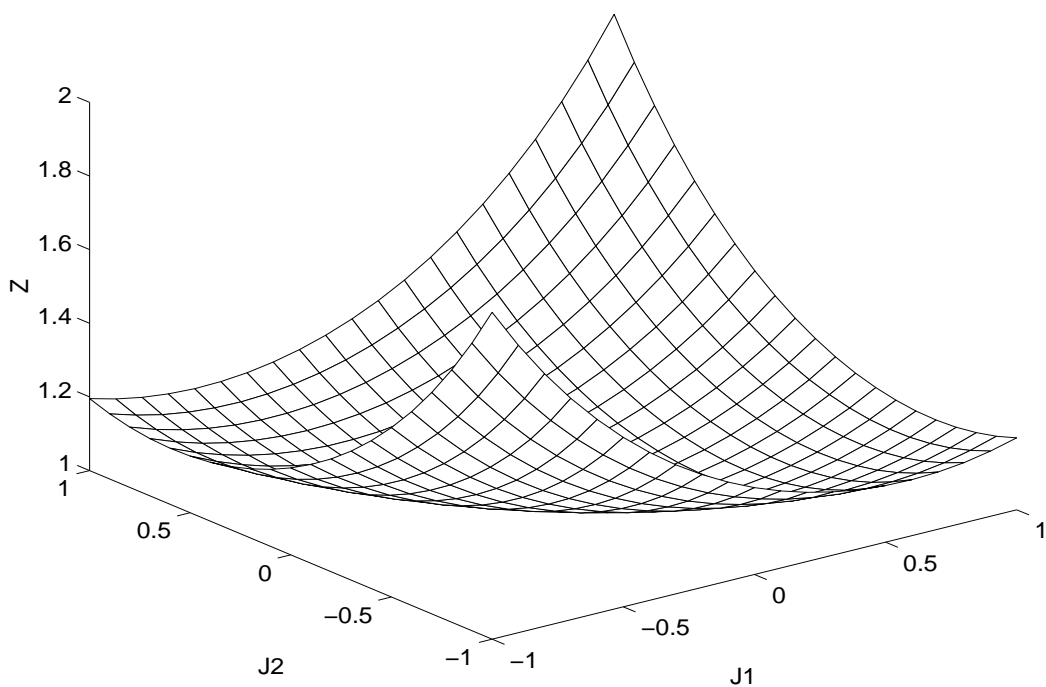,height=4in,width=5.5in}}
\caption{Galerkin Solution for M=1.0 and g=0.5}
\end{figure}

\clearpage
\begin{figure}
\centerline{\psfig{figure=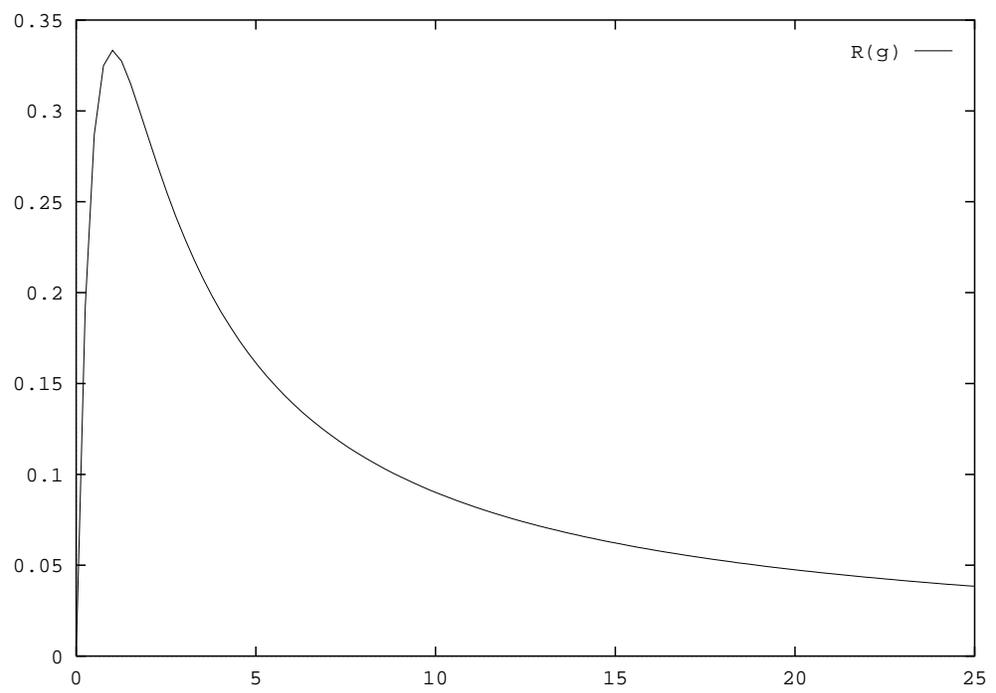,height=4.0in}}
\caption{Residual as a function of the coupling for 4th order truncation}
\label{gres}
\end{figure}

\clearpage

\noindent
inner product of the $R$ with a test function,
we are forming a linear combination of the SD equations organized
by $\epsilon$.  In other words, each Galerkin equation represents a
different
way of averaging the SD equations.  This is how the problem
of overdetermined equations is solved.  The galerkin coefficients
are not exact solutions, but averaged ones.  Taking the limit
$\epsilon \ra 0$ forces the error due to the averaging into
the higher order SD equations leaving the lower order ones
exactly satisfied.  Physically, taking this limit is reasonable
since we are only interested in the behavior of $Z$ at the
origin; physical Green's functions are derivatives of $Z$
at $J=0$.

To see how well truncation approximates
the theory, we examine the behavior of the residual
as a function of the coupling.
Notice that the
the truncated SD equations had exact solutions in
two limits, namely the free field limit and the infinite
coupling limit where the theory breaks up into a collection
of decoupled single sites.
Recall that exact solutions
corresponds to a vanishing residual.  For the two site
model truncated at 4th order, we have
the following
expression for the residual after the $\epsilon \ra 0$
limit has been taken.
\bg
R = \frac{27 G}{105 G^{2}  + 1033 G + 550} \; J_1^2  J_2
  - \frac{35 G}{105 G^{2}  + 1033 G + 550} \; J_2^3
\ed
where the coefficients of the residual measure
the error in satisfying the individual SD equations.

Graphing the nonzero coefficients versus
the coupling gives the expected behavior.
At zero coupling, the error is zero, and rises to
some maximum value as $g$ increases.
Asymptotically, it falls off as $g$ approaches
the strong coupling limit.
While the magnitude of the error decreases
at higher truncation order,
the value of $g$ that gives the maximum error
for a given truncation can vary
and depends on the details of the system.

To examine convergence, we truncate at higher orders.
The behavior of the residual
is shown in the plots for the two site model where
the $\epsilon \ra 0$ limit has already been taken.
We see that at higher orders $R$ becomes flatter
about $J=0$ which is in the center of the plot.
This indicates a reduction in the $L_{2}$ error which is
measured as the integral of $R$ about the origin.
\bg
\parallel R \parallel_2 = [ \,
\int_{ -\epsilon}^{\epsilon} R^2 \, d\vec{J} \,]^{1/2}
\ed
The flatter $R$, the smaller the integral of $R$ becomes.

\clearpage
\begin{figure}
\centerline{\psfig{figure=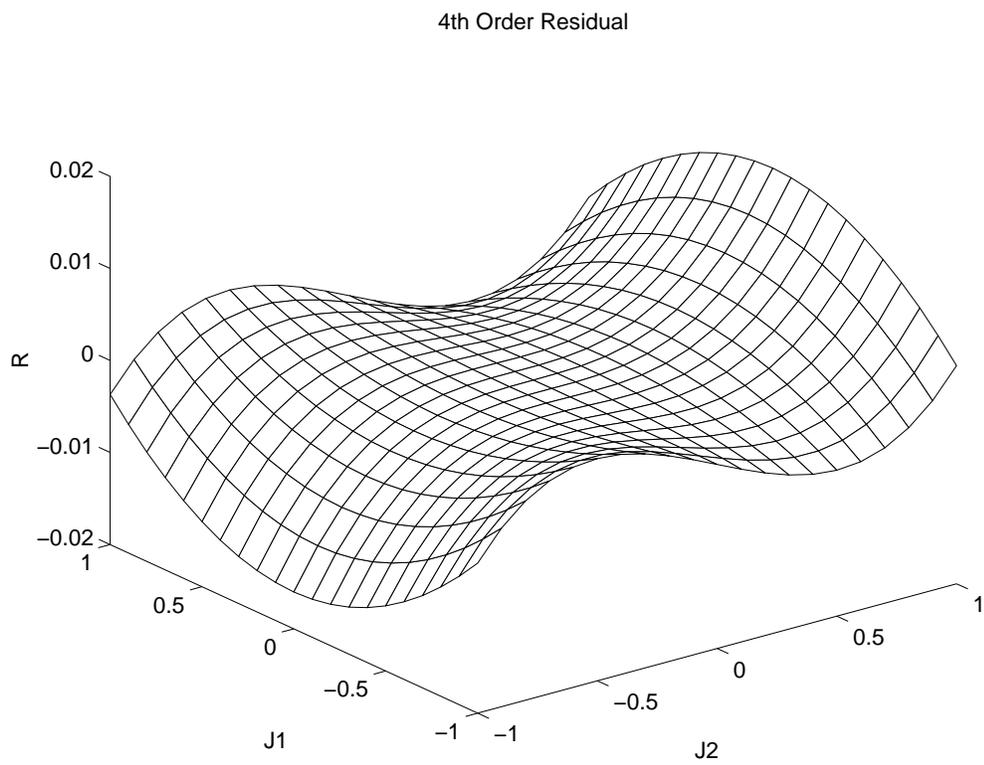,height=4in}}
\caption{Error Profile for 4th order truncation}
\label{res24}
\end{figure}

\clearpage
\begin{figure}
\centerline{\psfig{figure=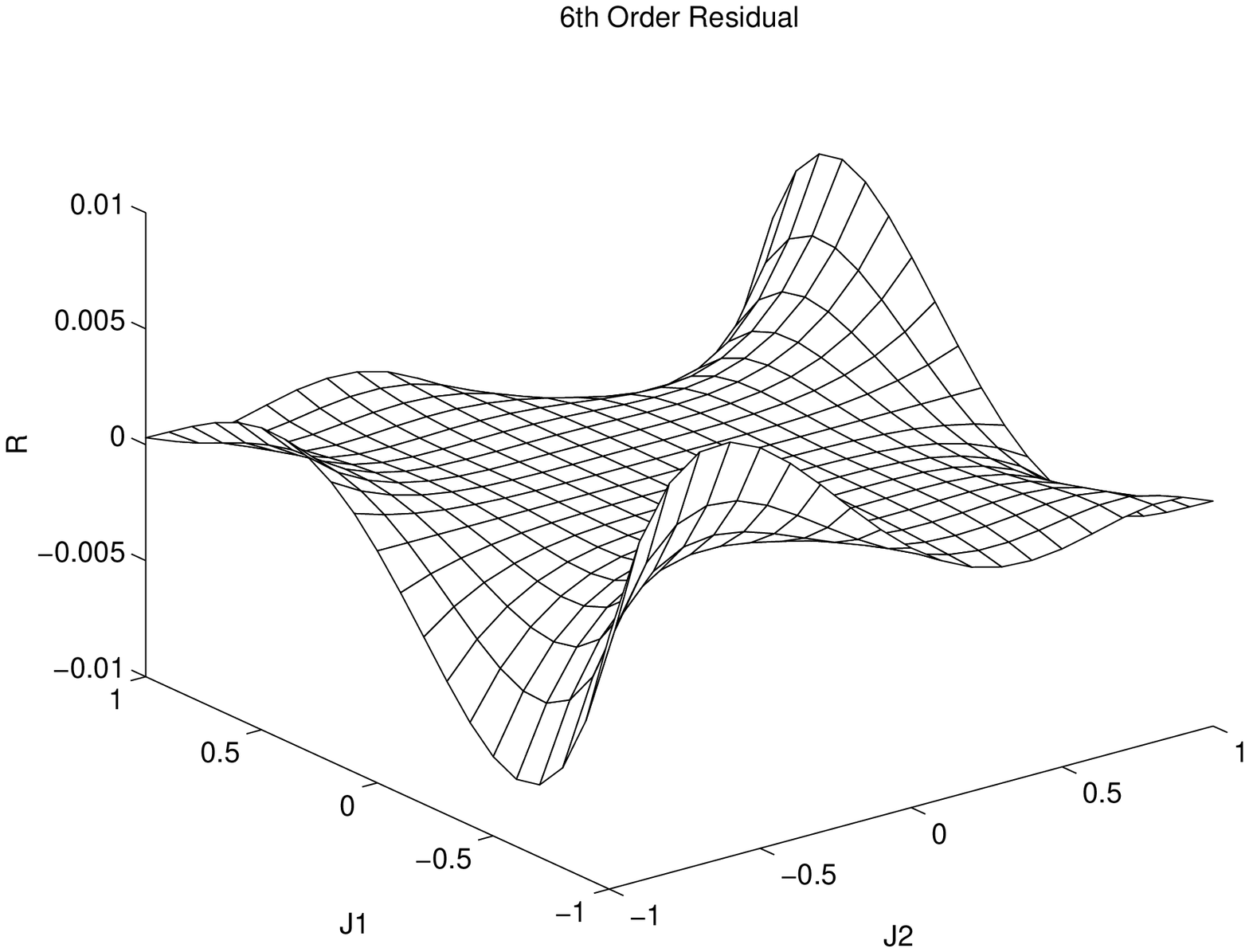,height=4in}}
\caption{Error Profile for 6th order truncation}
\label{res26}
\end{figure}

\clearpage
\begin{figure}
\centerline{\psfig{figure=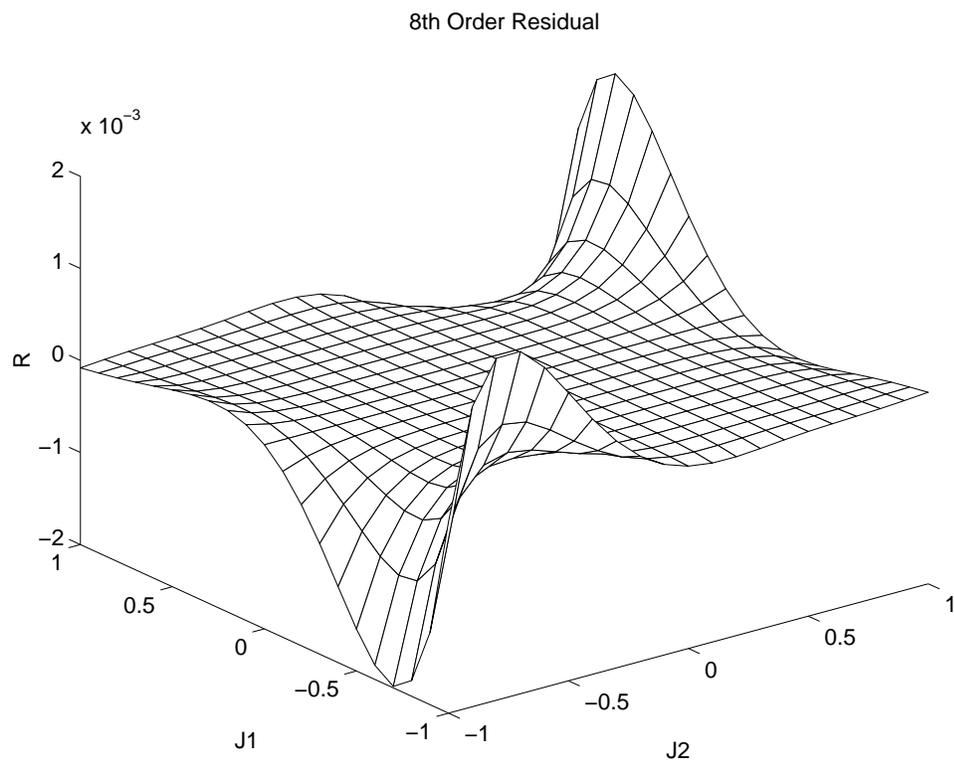,height=4in}}
\caption{Error Profile for 8th order truncation}
\label{res28}
\end{figure}
\clearpage

It is important to note that
the $L_{2}$ norm of $R$ is an extremely powerful tool for controlling
and measuring
the error due to approximations.  It is a general statement,
independent of the choice of basis functions, of how well a given
trial solutions approximates the differential equations.
Specifically, it tells us how far we are from the exact solution.
This is very much unlike other
methods.  For example, statistical errors can be calculated for Monte Carlo
simulations, but in general, systematic errors can be difficult to
quantify.  Similarly, in variational calculations, since you never
know the exact value of the ground state energy,
in principle,
you never know how close a given ansatz is to the exact ground
state. This must often be inferred from other methods or
from experiment.

\section{Lattice Invariant Polynomials}

In order to implement this method for larger systems,
we need a systematic way to construct polynomials
in the source variables that reflect the symmetries of our theory.
A general term in our power series
solution is represented by
\bg
a_{\{k1...kN\}} \; J^{k1}_{1} \ldots J^{kN}_{N}
\ed
where the subscripts index the sites at which the sources live and
the $k_{i}$ are non-negative integer exponents.
The coefficients $a_{\{k1...kN\}}$ are the Green's functions
of the theory and are required to be
translation invariant.
Therefore, we construct $Z_{T}$ to have the form
\bg
Z_{T} = \sum_{n,m}
{a}_{n,m}{P}_{m,n}
\label{expan1}
\ed
where the $P_{m,n}$ transform into themselves under
lattice symmetry operations.
They are the invariant polynomials in the sources of order $n$
with $m$ invariant classes for a given order.  The number of
invariant classes for a given order depends on both the
number of lattice sites and on the number of symmetry operations
for a given lattice.  For example, higher dimensional lattices
have larger symmetry groups and will therefore have fewer independent
invariants.
These polynomials
form a lattice invariant basis upon which we construct
solutions for  $Z$.
In one-dimension, lattice field theories with periodic boundary
conditions are invariant under translations and reflections of
the underlying lattice.  Such a lattice with $N$ sites is equivalent to
an $N$-sided polygon, and the lattice symmetry group corresponds
to the group of motions that transforms the polygon into itself.
These are the dihedral groups $D_{N}$ of order $N$.

     To construct polynomials invariant under $D_{N}$, we use
the following shorthand \cite{garcia} for the terms of our series
\bg
J_{1}^{k_{1}} \ldots J_{N}^{k_{N}}  \rightarrow  (k_{1} \ldots k_{N})
\ed
where $k_{i}$ is understood as the exponent associated with $J_{i}$.
In this notion, performing symmetry operations is straightforward.
Translations correspond to cyclic permutations of the exponents
\bg
(k_1 \hs \ldots \hs k_N)  \rightarrow (k_N \hs k_1 \hs \ldots \hs k_{N-1})
 \rightarrow \ldots
\ed
and reflections appear as inversions
\bg
(k_1 \; \ldots \; k_N) \; \rightarrow \; (k_N \; \ldots \; k_1)
\ed
Given any single term, you can generate its entire associated invariant
polynomial by applying the set of symmetry operations.

     Now we present an algorithm to systematically
generate all the invariant polynomials of a given degree.
We try to identify an independent
generator for each symmetry class. From these, we
can construct the full
polynomial $Z_{T}$ sorted into its invariant classes.
Identifying these independent generators is a
nontrivial problem.  One method could involve forming all
possible combinations of the $J_{i}$ at a given order, and
then sorting the full set into its
lattice invariant subsets.  A
``sieve algorithm" \cite{garcia} can be constructed that establishes a
criterion for sifting the independent representatives out of the full set.

    An alternate approach is to construct the generators directly,
without sorting through the full polynomial set.  One recurring problem
for all algorithms
is that duplicate terms can be generated when terms
have degenerate values for the $k_i$.
To avoid this, some sorting is usually
necessary.  For the present algorithm, sorting is needed only for a
restricted part of the full polynomial set.
Each term in $Z_{T}$ for a given order $n$ and for $N$ sites
is expressed as
\bg
( k1 \hs k2 \hs \ldots \hs kN )
\ed
This denotes some arrangement of the partitions of the integer $n$,
distributed among the $k_i$.  The other $N-n$
values are set to zero.  The partition
of an integer is defined as
\bg
a_{1}  +  a_{2}  +  ...  +  a_{n}  =  n
\ed
with
\bg
a_{1} > a_{2} > \ldots > a_{n} > 0.
\ed
Using this definition you can produce a hierarchy
of polynomial terms like
\begin{center}
\begin{tabular}{cccccccc}
               (&$n$   & 0 & 0 & 0 & & $\ldots$ & 0)  \\
               (&$n-1$ & 1 & 0 & 0 & & $\ldots$ & 0)  \\
               (&$n-2$ & 2 & 0 & 0 & & $\ldots$ & 0)  \\
               (&$n-2$ & 1 & 1 & 0 & & $\ldots$ & 0)  \\
                & \vdots & & & & & &           \\
               (&1   & 1 & $\ldots$ & 1 & 0 & $\ldots$ & 0)
\end{tabular}
\end{center}
Taking all possible permutations of these partition terms
gives all the terms in the expansion for $Z$ at order $n$.
In fact, this also generates many duplicate
terms.
Avoiding duplicates is one of the
principle difficulties in constructing any algorithm.

Next, we "bubble" the above partitions.
This means permuting all members {\em except} the first while preserving the
order of the nonzero elements.
\begin{center}
\begin{tabular}{cccccccc}
               (&$n-1$ & 1 & 0 & 0 & & $\ldots$ & 0)  \\
               (&$n-1$ & 0 & 1 & 0 & & $\ldots$ & 0)  \\
                & \vdots & & & & & &           \\
               (&$n-1$ & 0 & 0 & 0 & & $\ldots$ & 1)  \\
\end{tabular}
\end{center}
The word ``bubble" suggests that you insert zeroes among the nonzero
elements as if blowing bubbles between them.
This bubbling procedure while keeping the first element fixed
is the heart of the algorithm.  Remember we are trying to
construct generators directly.  This means we are looking for ways
to write down terms that are clearly {\em not} related by symmetry moves.
Since symmetry moves correspond to cycles and
inversions, fixing the first element
breaks these symmetries.  This guarantees that all of the
independent generators are contained in the set of bubbled partitions.

There are special cases when the set of bubbled partitions
does not contain all the generators. Since bubbling preserves the order
of the nonzero elements, this excludes terms that do not have the
descending hierarchy among the nonzero elements.  For example
on a four site lattice, taking
all permutations of (4 3 2 0) includes terms like (4 2 3 0) which
are missed under strict bubbling.  An easy remedy is whenever you
have two or more nondegenerate, nonzero elements, not including the first
one, permute these elements before bubbling.
In general, these permutations are only a minor part of the algorithm.
Also note that not all bubbled partitions
are independent.  This is because a remnant of our lattice symmetry
remains even after fixing the first element, namely reflections about
site 1; the one we fixed.  Therefore, we group
terms that are related by reflections.
After deleting the extra terms, the ones remaining are the independent
generators.

     Using these independent generators, you can recover their associated
invariant polynomials by applying symmetry moves.  Some of these terms,
may either transform into themselves or into terms already
present.  When building a individual symmetric polynomial, you must check for
duplicates.
Some of the terms may transform into
themselves or into terms already present in the individual polynomial.
Since the invariant polynomials are generally small - the
maximum size is equal to the total number of symmetry moves for the lattice
($2N$ for the dihedral group $D_{N}$),  the number of checks is small
and is done quickly.

     The above algorithm for identifying the generators of the $D_{N}$
invariant
polynomials generalizes immediately to the more complicated symmetry groups
associated with higher dimensional lattices.  In two-dimensions, there
is a source variable $J_{(i,j)}$ living on each site of a square lattice and
the power series must be invariant under all $2D$ symmetry moves.
\def\vs{\vspace{0.15in}}
\def\hs{\hspace{0.1in}}
\begin{center}
\begin{tabular}{ccc}
\vs
$J_{1,N}$ & $\ldots$ & $J_{N,N}$ \\
$\cdot$ & & $\cdot$ \\
$\cdot$ & & $\cdot$ \\
$J_{1,1}$ & $\ldots$ & $J_{N,1}$ \\
\end{tabular}
\end{center}
In analogy with the $1D$ case, we represent a general term in our
power series
\bg
J_{1,1}^{k_{(1,1)}} \hs  \ldots \hs J_{N,1}^{k_{(N,1)}} \hs \ldots \hs
J_{1,N}^{k_{(1,N)}} \hs \ldots \hs J_{N,N}^{k_{(N,N)}}
\ed
as a 2D array of the integer exponents
\[
\left| \begin{array}{ccc}
k_{(1,N)} & \ldots & k_{(N,N)} \\
\cdot & & \cdot \\
k_{(1,1)} & \ldots & k_{(N,1)}
\end{array}
\right| \]
Here again, performing symmetry moves is straightforward.  Translations with
periodic boundary conditions corresponds to cyclic permutations of
all rows or all columns.  Similarly, reflections
corresponds to inversions of all rows or all columns.
Euclidean rotations are obtained by
rotating the arrays by 90 degrees
while reflections about the diagonal are performed as products
of a rotation plus a reflection.

The algorithm we developed in the last section can be applied immediately.
For a given order, we perform the partitions of $n$.  If necessary,
we permute nonzero elements.  Then, we fill the 2D array from
left to right beginning at the bottom left corner.  Bubbling is performed
by fixing the lower left corner element and inserting zeroes such that
the nonzero elements are moved from left to right starting at the bottom
row and moving to the top successively.
As in the 1D case, even after fixing one element, there is still a remnant
lattice symmetry in the set of bubbled partitions, namely reflections
about the bottom left corner.
Once this symmetry has been distilled,
you are left with the independent generators.  The full polynomial can be
constructed by performing symmetry moves on these generators.  Again,
care must be taken in checking for duplicates within
each individual polynomial.
This procedure is identical for cubic and hypercubic lattices.
Eventhough these groups can be large,
they have been thoroughly studied
and their elements tabulated \cite{mandula}.

\section{Numerical Calculations}

Now, we look at a variety of solutions to lattice $\phi^4$
field theory.  Implementing the method numerically, we calculate
Green's functions and the corresponding mass-gaps for lattices
in $D=2,3,4$.  In $D=1$, reasonable large lattices can be examined.
Results for these systems are given in \cite{garcia},\cite{galerkin}.
In higher dimensions, the rapid increase in complexity
with the number of sites limits both
linear size of the systems and the truncation orders
available.  In $D=2$, $8\times8$ lattice models truncated at 4th
order are possible on a workstation.
But in $D=3,4$,

\clearpage

\begin{figure}
\centerline{\psfig{figure=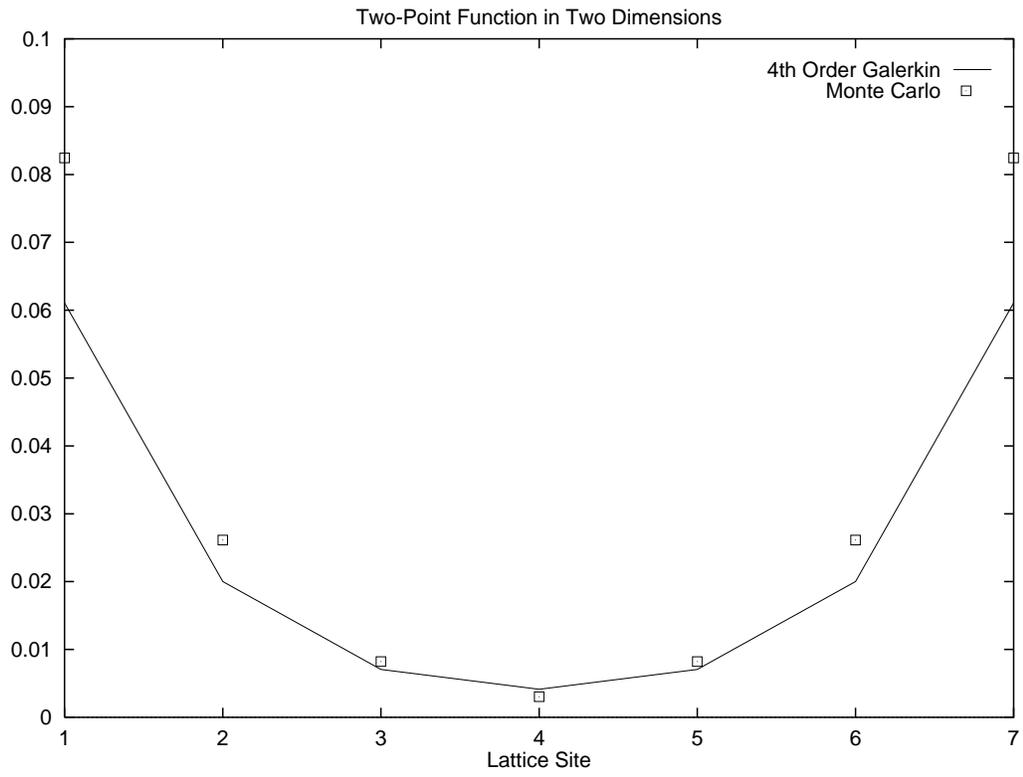,height=4in}}
\caption{Propagator for 8x8 lattice with M=1.0, g=0.5}
\end{figure}
\clearpage

\begin{figure}
\centerline{\psfig{figure=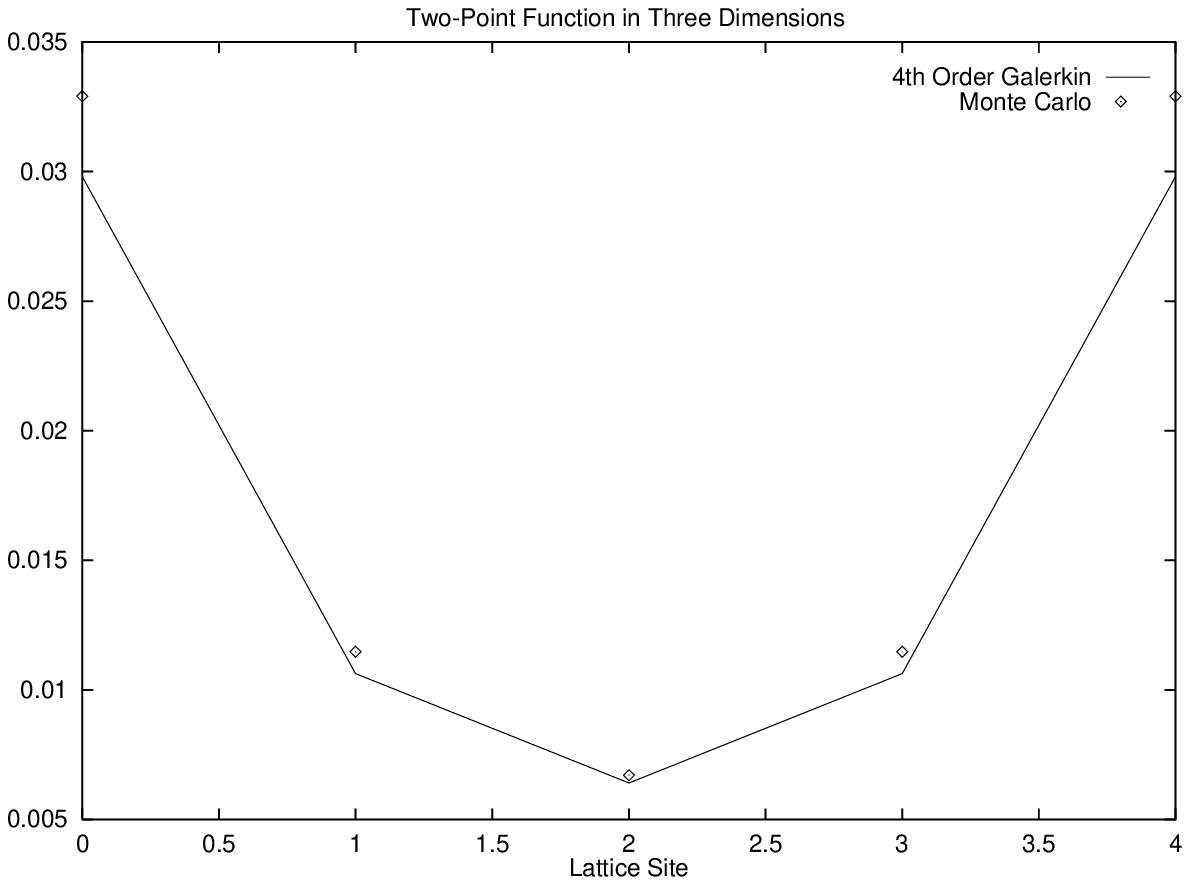,height=4in}}
\caption{Propagator for $4^3$ lattice with M=1.0, g=0.5}
\end{figure}
\clearpage

\begin{figure}
\centerline{\psfig{figure=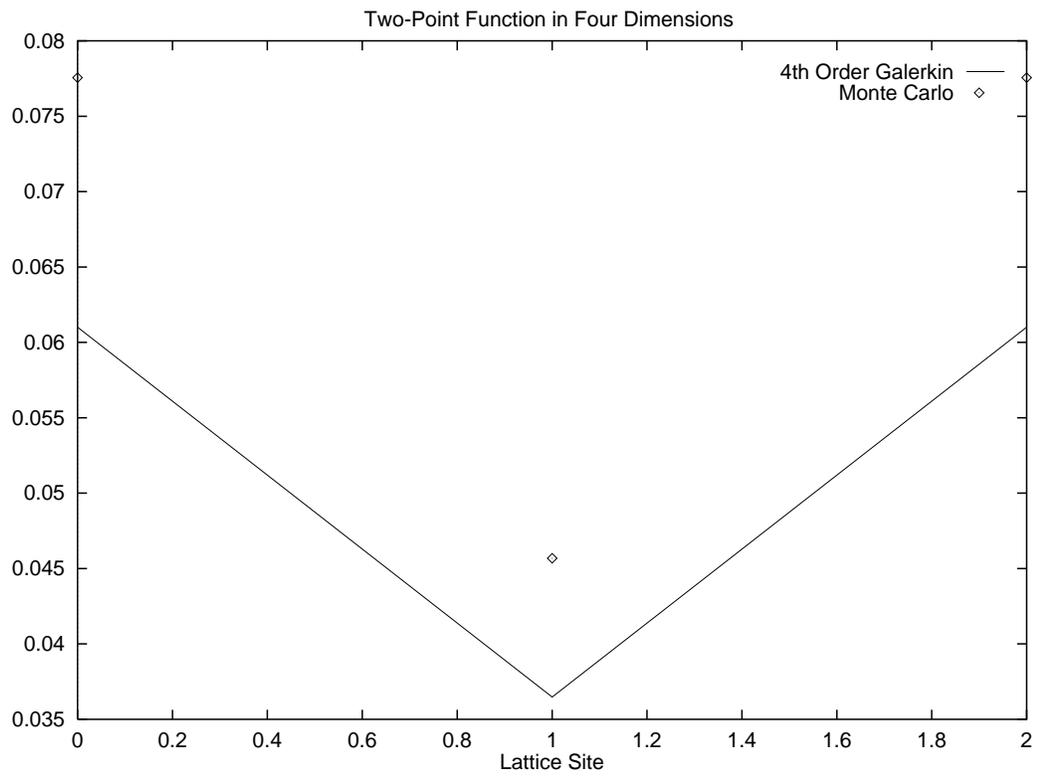,height=4in}}
\caption{Propagator for $2^4$ lattice with M=1.0, g=0.5}
\end{figure}
\clearpage

\noindent
lattices larger
than $4^D$ truncated at higher than 4th order are prohibitive.
For each calculation, we compare our results
against Monte Carlo simulations performed using the Metropolis algorithm.

Calculations with the Source Galerkin method are very clean
and efficient, especially when compared to Monte Carlo.
The complexity of calculations for polynomial basis functions
scales with the total number of sites and the truncation order.
It is independent of the dimensionality of the lattice and the form
of the interaction.  For polynomial interactions,
the differential
equations are always be linear.  The bulk of a calculation involves
a single matrix inversion for a given set of parameters.
This is
in contrast to Monte Carlo where many sweeps through the lattice
are necessary to reduce statistical error.  As the plots show,
the Source Galerkin calculations
show rapid convergence even at
intermediate couplings and using only a fourth order polynomial.

\section{An Alternative to the Galerkin Method}

We have emphasized the flexibility of the Source Galerkin method
with respect to choice of basis functions for our trial
solution $Z_T$.  Once we have a made a choice, we fit
the approximation to a solution of the functional
differential equations by using the Galerkin method.
It is important to note that there is nothing
fundamental about the Galerkin method.  Its
function is to minimize the error due approximation
and force convergence to the exact solution in
a controlled and systematic way.
There are many ways to accomplish this, the Galerkin
method being just one.  Other procedures might be
desirable based convenience.  Here we present one
such alternative.

Previously, we have emphasized that solving the lattice
functional equations with a truncated power series leads
to an overdetermined set of algebraic equations for the
expansion coefficients.
Because the equations cannot be solved exactly,
the Galerkin method is used to find an average solution.
In terms of the set of differential equations, the Galerkin
method gives a ``weak" solution where the error due to fitting
the approximate solution is averaged over a hypercube
centered at the origin of source space ranging from
$+\epsilon$ to $-\epsilon$ in each direction.  Since we
want an especially good fit at the origin $(J=0)$, after the
calculation we take $\epsilon \ra 0$.  In terms of the lattice
SD equations, the combination of the Galerkin method
and the $\epsilon \ra 0$ limit
gives a set of partially
satisfied relations.  For the 4th order truncated system, we had
more equations than unknowns, but after $\epsilon \ra 0$,
all SD equations except for those at the highest untruncated
order are solved exactly with the error being pushed
successively to higher orders as the truncation increases.

With these observations in mind, consider some of the practical
difficulties of the Galerkin method.  The bulk of the computational
effort involved in a Source Galerkin calculation is the inversion
of a potentially very large matrix.  In fact, the principle constraint
of this approach is the size $N$ of the matrix that you can invert.
Since for realistic systems these matrices must be inverted
numerically and since they can become very large, it is
critical that they be well-behaved.

For a numerical implementation of the method, this matrix is
usually badly conditioned.
Each row represents a different linear
combination of {\em all} the SD equations
where the Galerkin weightings depend heavily
on $\epsilon$.  But in order to perform
the numerical extrapolation, we choose values of $\epsilon$ that
are very small, typically $\epsilon \sim 1/10$.  After integrating
out the sources, the weightings can depend on a fairly high
power of these small numbers.
Since $\epsilon^{n}$ dominates over other numerical factors such as
the coefficients in the SD equations and the system
parameters $(M,g)$, variation between rows can be small.
Since the lattice
size and the truncation order is limited by the size of the matrix
that can be inverted, we ideally would like sparse,
well-conditioned matrices.
Powerful algorithms exist for inverting sparse matrices
that are not only efficient in terms of time and memory resources,
but allow inversion of significantly larger matrices and with
greater accuracy.

One possible method to accomplish this is to deal with the
SD equations directly, bypassing the functional
formulation and the Galerkin method.
We know from looking at the solution to the SD equations
that after using the Galerkin method and taking the limit
$\epsilon \ra 0$, all the low order equations are exactly
satisfied, some of the highest order equations are exactly
satisfied, and the remaining high order equations have
some sort of average solution.  It seems natural to ask why not
solve the SD equations directly, and forget about power series,
and integrating out the sources, and numerical extrapolations.
Plus we obtain the practical advantage of sparse matrices.

In addition, this procedure considerably simplifies the fermion
problem.  In its original form, the Source Galerkin
method is relative symmetric in its treatment of bosons and
fermions, the only difference being
that fermionic sources anticommute.
While on the surface, this does not seem to be a problem, it
does create a obstacle for constructing Galerkin solutions
to the Grassmann differential equations.  This problem must
be solved by defining a modified
form of Grassmann integration \cite{fermion}.
Using a scheme that attempts to solve the
fermionic SD equations directly would be free not only of the
practical handicaps of the boson method, but also would not
require this modified integration.

We return to the two site model to examine these ideas.
As a first attempt to solve the overdetermined equations,
we notice that only one coefficient,$a_4$, in the set
$\{a_1, \ldots, a_5 \}$ is constrained by more than one equation,
namely the fourth and the fifth
\bg
\alpha \, a_{4} - 2 a_{5} - 2 a_{2} & = & 0.0247 \no \\
\alpha \, a_{4} - 2 a_{3} & = & - 0.0961.
\ed
To have a coefficient be constrained by an equation, we mean
that the lattice KG operator acts directly on it in a particular
SD equation.  This operation is signaled by the factor of $\alpha$
multipling the coefficient.  Also notice that the Source Galerkin
method has pushed {\em all} the error into these two relations
as indicated by the nonzero RHS.  Furthermore, remember that
for this model
\bg
a_{4} & = & \langle \phi_{1}^{3} \phi_{2} \rangle
= \langle \phi_{1} \phi_{2}^{3} \rangle. \no
\ed
We have mentioned that the truncated SD equations are overdetermined.
This is due to a symmetry mismatch
between the lattice symmetric $Z$
and the individual differential source
operators $\hat{D}_{i}$ which are symmetric
only {\em about} site $i$.  The individual $\hat{D}_{i}$ are not
translation invariant.  For the two site model, we use the single
differential equation to constrain $Z$
\bg
\hat{D}_{i=1} \, Z = 0.
\ed
Clearly, $a_4$, as defined above is not invariant under the symmetry
group of $\hat{D}_{i=1}$. It is this tension that causes the additional
equation for $a_4$ to be generated, and leave the set overdetermined.

We can, therefore, ask, if there are two equations for one unknown
and both relations are only partially satisfied, then why not put
{\em all} the error into one equation and solve the other
one exactly.
In other words, we solve a
carefully chosen {\em subset} of the SD equations such that there
is only one equation for each unknown.
This is trivial way to calculate.  We know what
the lattice SD equations are.  For example,
two point function obeys
\bg
[ \, 2 - ( d^{+}_{i} + d^{-}_{i} ) + m^2 \, ] \, G_2(i,j)
+ g \, G_4(i,i,i,j) = \delta_{i,j}.
\ed
Higher order equations can be found easily.
In large scale problems, the matrix becomes
extremely sparse with no more that a few nonzero elements per row.
Generation of the unknown Green's functions
is a simple exercise in combinatorics and symmetry identical to
constructing invariant polynomials.
Thus, we have found an approach with no sources, no integrals,
and no $\epsilon$.  In the end, we have a sparse matrix to invert.

\begin{figure}
\centerline{\psfig{figure=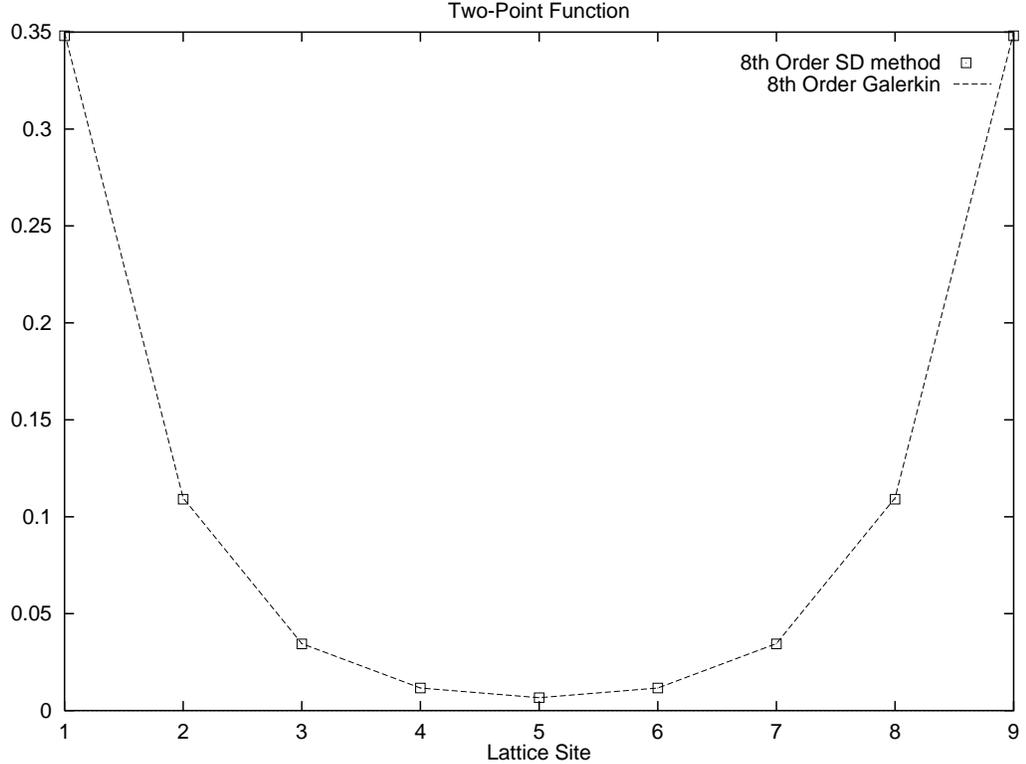,height=4in}}
\caption{Comparison of SD method with integral Galerkin with M=1.0, g=0.5}
\end{figure}
\begin{table}
\begin{center}
\begin{tabular}{|c|c|c|c|c|c|c|c|} \hline
\multicolumn{1}{|c|}{$|i-j|$} &
\multicolumn{2}{|c|}{4th order} &
\multicolumn{2}{|c|}{6th order} &
\multicolumn{2}{|c|}{8th order} &
\multicolumn{1}{|c|}{Monte Carlo} \\ \hline
  &  SD    & Galerkin &   SD   & Galerkin &   SD   & Galerkin &        \\
\hline
0 & 0.3274 & 0.3268   & 0.3592 & 0.3601   & 0.3481 & 0.3497   & 0.3426 \\
\hline
1 & 0.0957 & 0.0948   & 0.1167 & 0.1177   & 0.1090 & 0.1096   & 0.1087 \\
\hline
2 & 0.0278 & 0.0271   & 0.0383 & 0.0392   & 0.0344 & 0.0345   & 0.0329 \\
\hline
3 & 0.0086 & 0.0081   & 0.0136 & 0.0143   & 0.0117 & 0.0117   & 0.0106 \\
\hline
4 & 0.0045 & 0.0042   & 0.0080 & 0.0086   & 0.0066 & 0.0063   & 0.0057 \\
\hline
mass-gap & 1.2410 & 1.2647 & 1.1240 & 1.1189 & 1.1630 & 1.1649 & 1.1190
\\\hline
unknowns &   34   &   34   &  160   &   160  &  600   &  600   &  \\\hline
\end{tabular}
\end{center}
\caption{Comparison of SD, Galerkin, and MC for 8 site lattice}
\end{table}

We performed several calculations on 10 site lattices,
truncating at successive orders and comparing with both
Monte Carlo and the integral Galerkin method.  We see from
the graph and the table that there is good agreement between the
three approaches.
Despite
these results, it is uncertain if this approach will have problems
for large systems at very high order.
The reason is that by solving only
a subset of the SD equations, we ignore more and more
equations, the additional inconsistent ones, as we go to
higher order.  But these equations are relations that must
be obeyed by the exact solution to the differential
equations.  By discarding, these equations, we have
introduced an uncontrolled approximation.  Ignoring these
low order equations means that there is no
guarantee that
we will converge to the exact solution.
As a practical matter, though, we rarely truncate
above 4th or 6th order, so that this loss of a ``rigorous"
notion of convergence might not be a real problem, and in
fact, may be a small price for the
simplicity of this approach. For many examples,
it has successfully captured many of the features of the truncated theory.
In the next section, we outline how to
define a rigorous notion of convergence in the lattice Green's
function approach.
It will allow us to control the approximations
in much the same way as the Galerkin method.

\section{Flatness Criterion for Controlled Convergence}

An important component of the Galerkin method is that it gives
a rigorous, well-defined way to control the error due to the
truncation approximation.  At each level of truncation, Galerkin
smoothed out the error, and by
considering a sequence of successive truncations,
we can improve our approximations and have a well-defined notion
of convergence to the solution of the differential equations.
Since we are proposing
to dispense with at least the external trappings of the Galerkin
method, we need to formulate a new way to control our approximations
and enforce convergence to the correct solutions.

Recall that the Galerkin method, and in
fact most spectral methods, are based
on a trial solution to some differential equation.  Since
an exact solution is usually not available, some error
is always present.  Various approaches differ principally in
how they deal with the problem of controlling and minimizing
this error.  The Galerkin method, for example, tries to
minimize the area under the error function $R$.  Effectively,
it replaced the condition $R=0$ pointwise throughout the domain,
which is only true for the exact solution, with $\int_{D} R \, [dx]=0$
i.e. the error vanishes on average across the domain.
For problems in quantum field theory,
we are only interested in minimization at
the origin, so it is only the area of $R$ infinitesimally close
to $J=0$ that is relevant.  The question we ask is - are there
other ways to enforce this condition of minimized area about
zero without actually calculating integrals?

In the previous section, we developed an approach
to find Green's functions by solving a subset of the lattice
SD equations.  As mentioned, this introduces an uncontrolled
approximation.  To remedy this we would like to deal with the
full set of equations.
Of course, we know that the truncated SD equations
can only be solved on average since they are overdetermined.
If we can find some criterion to define what an average solution
means in this SD context, then maybe we can retain the simplicity
of the Green's function approach, and in addition
gain the rigorous notion of convergence
that characterizes the Galerkin method.
Towards this, first
notice that the residual has the form of a Taylor's series centered
at the origin where the derivative coefficients are the SD equations
which we would like to deal with directly.
\bg
R = \frac{dR}{dJ_1} |_{J=0} \, J_1
  + \frac{dR}{dJ_2} |_{J=0} \, J_2
  + \frac{d^3R}{dJ_1^3} |_{J=0} \, J_1
  + \frac{d^3R}{dJ_1^2 dJ_2} |_{J=0} \, J_1^2 J_2
  + \ldots
\ed
Obviously, the ``flatter" $R$ is about the origin (i.e. has vanishing
low order derivatives), the less area it will subtend.  In fact,
this is effectively what the Galerkin method does after $\epsilon \ra 0$
as can be seen from the graphs of the two site residuals.  At
successively higher truncations, $R$ becomes flatter at the origin.

\begin{figure}
\centerline{\psfig{figure=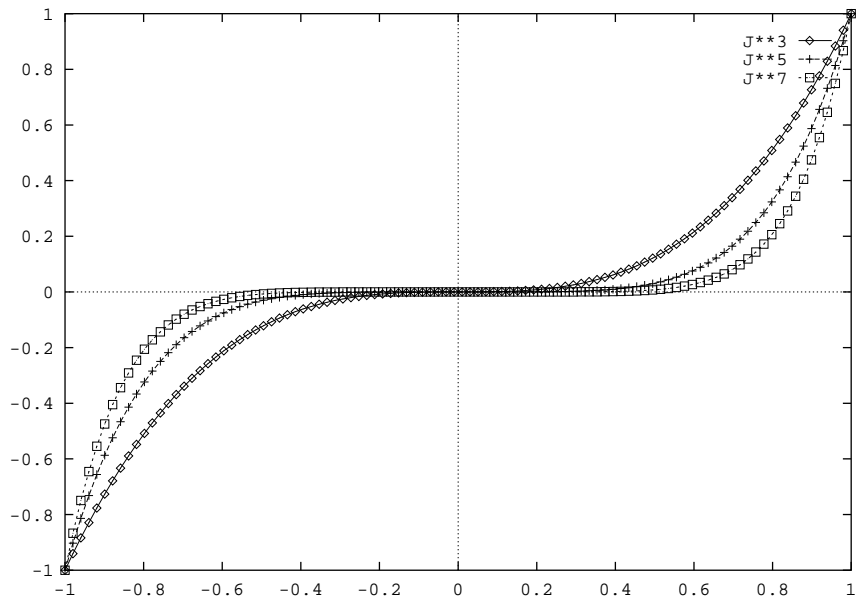,height=3.5in}}
\caption{Flattening of the residual with higher order}
\end{figure}

We propose to control the error by invoking a ``flatness
criterion" to replace minimization of integrals.  Of course, the
flattest function that goes through the origin is $R=0$, but we
know that $R$ cannot vanish because we do not have an exact solution.
$R$ is necessarily a nontrivial function of the sources.  Therefore,
we zero as many of the low order derivatives of $R$ as possible,
and then mix only the highest order derivatives in order to reduce
the number of equations.  Flattening $R$ about $J=0$ has exactly
the same error content as minimizing the area and taking the $\epsilon \ra 0$,
only there are no integrals and no numerical extrapolations.
We can make this method systematic by truncating $Z$ at successively
higher orders and zeroing all derivatives except those at the highest
untruncated order.  As $R$ becomes flatter and flatter about $J=0$,
the error is pushed into successively higher derivatives and $R$
smoothly approaches zero as the truncation goes to infinity.

Notice that to construct a notion of convergence and averaging,
we appealed to $Z$ and the functional formulation to provide
mathematical structure.  But in practice, you never need to
deal with $Z$ or with sources at all.  The SD equations can
be constructed quickly and independently of the partition function.
It is even easy to determine what derivative of $R$ a
particular equation corresponds to eventhough we never calculate $R$.

To obtain a complete theory, we must determine
which equations to mix at the highest order and with what weightings,
if any, to assign them.  To do this, we can look for clues in
the integral Galerkin method.  In that approach, inner products
of $R$ are taken with linearly independent test functions.  These
test functions act as projectors in the Galerkin function space
and their precise functional form determines how the SD equations
are mixed, and which equations become exact (or not) when
$\epsilon \ra 0$.  In principle, there should be a map
between a particular choice of test functions $\{T_i\}$ and
how the SD equations are averaged.  This is a nontrivial task
because these rules are coded not only into the form of the
inner product and the set of test functions, but also depend
on the $\epsilon \ra 0$ limit.
At present only the loose
form of these rules is known.
This problem is
being pursued and its resolution will be presented elsewhere.

\section{Conclusions}

In this paper, we have continued work on a new numerical method
for quantum field theory called the Source Galerkin method.
It is based on the differential formulation of quantum
field theory in the
presence of an external source.
By examining the functional differential equations for a theory
on a finite lattice, we obtain a set of coupled
partial differential equations for the
generating functional $Z$.  For nonlinear field theories with
polynomial interactions, the equations to solve are always linear.
Once we have obtained $Z$, we can extract the Green's
functions by functional differentiation.

To construct solutions, we can expand $Z$
on {\em any} complete set of functions
in the source variables $\{J\}$.  A particularly simple choice is
polynomial functions.  We saw that these functions gave very
rapid convergence even using low order polynomials.
Calculations were efficient to perform and produced very clean
numbers.
The bulk of any calculation involved only a single matrix inversion.
Due to computational complexity, polynomial basis
functions are limited to small systems.
A more general approach is required for large systems.  We
will have more to say about this in future communications.

Because our solutions for $Z$ are necessarily approximate, we
found the Galerkin method very powerful for controlling error.
It will fit any trial solution to a solution of the differential
equations.  This can be formulated in a systematic way, guaranteeing
convergence to the exact solution.  The residual function was
especially useful for quantifying and controlling the error
due to approximations.  It is a very direct and precise measure
of how well our approximations fit to the differential equations.
This strong control of error should be contrasted against other
methods especially Monte Carlo and variational techniques
which rely on less precise determinations.

In another paper, we will show that the Source Galerkin method
is especially powerful for fermion systems \cite{fermion}.  The fermionic
formulation is identical except that the sources anticommute.
Because this method is deterministic and allows for systematic
approximations, it is very useful for examining these
systems.

\section{Acknowledgements}

We would like to thank Stephen Hahn for useful discussions
and for writing code used for the calculations in Section 6.
In addition, we are indebted to Santiago Garcia for many discussions.
GSG would like to thank Vance Faber for the Hospitality of C-3 at Los
Alamos National Laboratory where some of this work was done.

\end{document}